\documentclass[]{raa}  
\usepackage{amssymb, amsmath}	
\usepackage{newtxtext,newtxmath}
\usepackage[pagebackref=true]{hyperref}
\usepackage{graphicx}	
\usepackage{verbatim}
\usepackage{natbib}
\usepackage{pdflscape}
\usepackage{float}
\bibpunct{(}{)}{;}{a}{}{,}
\begin{document}
\title{Relativistic Low Angular Momentum Advective Flows onto Black Hole and Associated Observational Signatures}

\volnopage{Vol.0 (20xx) No.0, 000--000}      
   \setcounter{page}{1}          

\author{Jun-Xiang Huang \inst{1}  
\and Chandra B. Singh  \inst{1}}  
 \institute{
 South-Western Institute for Astronomy Research, Kunming 65050, China; {\it wacmkxiaoyi@gmail.com, chandrasingh@ynu.edu.cn}\\
 \vs\no
   {\small Received 20xx month day; accepted 20xx month day}}

\abstract{
We present simulation results examining the presence and behavior of standing shocks in zero-energy low angular momentum advective accretion flows and explore their (in)stabilities properties taking into account various specific angular momentum, $\lambda_0$. Within the range $10-50R_g$ (where $R_g$ denotes the Schwarzschild radius), shocks are discernible for $\lambda_0\geq 1.75$. In the special relativistic hydrodynamic (RHD) simulation when $\lambda_0 = 1.80$, we find the merger of two shocks resulted in a dramatic increase in luminosity. We present the impact of external and internal flow collisions from the funnel region on luminosity. Notably, oscillatory behavior characterizes shocks within $1.70 \leq \lambda_0 \leq 1.80$. Using free-free emission as a proxy for analysis, we shows that the luminosity oscillations between frequencies of $0.1-10$ Hz for $\lambda_0$ range $1.7 \leq \lambda_0 \leq 1.80$. These findings offer insights into quasi-periodic oscillations emissions from certain black hole X-ray binaries, exemplified by GX 339-4. Furthermore, for the supermassive black hole at the Milky Way's center, Sgr A*, oscillation frequencies between $10^{-6}$ and $10^{-5}$ Hz were observed. This frequency range, translating to one cycle every few days, aligns with observational data from the X-ray telescopes such as Chandra, Swift, and XMM-Newton.
}

\authorrunning{J.-X. Huang \& C. B. Singh}            
   \titlerunning{Low Angular Momentum Flows and Observations}  
\maketitle



\section{Introduction}
\label{sec:1}
The existence of black holes has been firmly established through groundbreaking observations, notably of M87*, as presented by \citet{2019ApJ...875L...1E}. M87* is a supermassive black hole  of $10^9\text{ M}_\odot$ residing at the heart of the M87 galaxy. The M87 galaxy has been the subject of numerous observational studies, with many focusing on flux variability in recent times. For instance, \citet{2021RNAAS...5..136H} investigated the ultraluminous X-ray flaring sources within the M87 cluster, while \citet{2021ApJ...919..110I} studied the X-ray flux variability emanating from M87* itself. In contrast to M87*'s immense mass, the supermassive black hole at the center of our own galaxy, Sgr A*, has a considerably lower mass of $4\times10^6\text{ M}_\odot$, as highlighted by the observations by \citet{2022ApJ...930L..12E}. It is crucial to remember that black holes themselves are not luminous and cannot be directly observed. As such, when observing these celestial phenomena, the luminosity and contributions of the surrounding accretion disk must be duly considered.

\citet{1973A&A....24..337S} introduced the standard thin disk model, abbreviated as SSD, which however, seems misaligned with the observations of Sgr A*. Specifically, the observed luminosity of Sgr A* falls short of SSD predictions by a factor of five orders of magnitude. Additionally, the spectral characteristics of Sgr A* diverge markedly from those anticipated by the SSD model. In response to these discrepancies, the scientific community explored advective accretion models. Pioneered by \cite{1952MNRAS.112..195B}, the zero-angular momentum model emerged, followed by the advection-dominated accretion flows (ADAFs) with high-angular momentum, as highlighted by \citep{1994ApJ...428L..13N}. Subsequent studies \citep[e.g.,][]{2008NewAR..51..733N, 2014ARA&A..52..529Y} affirmed the validity of these advective accretion models, indicating their capability to more accurately describe the observed characteristics of Sgr A* \citep[e.g.,][] {2012ApJ...761..129Y, 2013ApJ...767..105L}. Multiwavelength investigations into Sgr A* have discerned two distinct states: a quiescent phase and a flaring phase \citep[e.g.,][]{RevModPhys.82.3121}. Empirical observations reveal that X-ray and infra-red (IR) flares typically last between 1-3 hours and manifest several times in a day. Notably, the variability in X-ray flux exhibits a pronounced shift, spanning more than two orders of magnitude when transitioning to the quiescent state \citep[]{2017MNRAS.468.2447P}.

Magnetohydrodynamic (MHD) investigations into magnetized disks have consistently highlighted the pivotal role of magnetic fields in the evolution of accretion disks \citep[e.g.,][]{2000ApJ...532L..67M, 2001PASJ...53L...1M, 2001MNRAS.322..461S, Igumenshchev_2003, 2003PASJ...55L..69N, 2012MNRAS.426.3241N, Yuan_2012, Yuan_2015}. A subset of these numerical MHD simulations has specifically delved into the mechanisms underpinning the flares of Sgr A* \citep[e.g.,][]{Chan_2009, Dexter_2009, Dodds-Eden_2010, Ball_2016, 2017MNRAS.467.3604R}. In a paradigm that factors in general relativity within the MHD framework (GRMHD), \cite{2017MNRAS.467.3604R} unveiled emission patterns characteristic of thermal electrons. Meanwhile, other studies \citep[e.g.,][]{2021MNRAS.507..983L, 2022arXiv221010053K, 2023MNRAS.518.1656M, 2023ApJ...944L..48L} have spotlighted the Lense-Thirring effect within accretion disks, leading to the emergence of high-frequency quasi-periodic oscillations (HFQPOs). Further elucidating these phenomena, \cite{Ball_2016} attributed rapid X-ray variability to electrons being accelerated via magnetic reconnection within areas exhibiting high magnetization in proximity to black holes. Additionally, \cite{2017MNRAS.468.2552L} proposed a scenario aiming to decode the X-ray flares of Sgr A* using the framework of episodic mass ejection. While the aforementioned research largely leans into the paradigm of high angular momentum flow, culminating in luminosity flares of Sgr A* spanning 1-3 hours, long-standing observations spanning nearly two decades—from telescopes such as Chandra, Swift, and XMM-Newton—reveal a pattern of luminosity variability extending across several days.

On the other hand, models characterized by low angular momentum advective flows are observed to support stable/unstable standing shocks and positioned between the flow models: advection-dominated accretion flow (ADAF) (where the specific angular momentum of the flow is some fraction of the Keplerian value at a given radius) and Bondi accretion flow (where the specific angular momentum is zero),\citep[e.g.,][]{1996ApJ...464..664C, 2006MNRAS.370..219M}. In the present study, the flow with low angular momentum close to a non-rotating black hole is usually around the corresponding marginally stable value ($1.837 R_{g}c$) where $R_{g}$ and $c$ are Schwarzschild radius and speed of light respectively. These shocks in such flows are postulated to induce luminosity variability. The simulation works by \cite{Dongsu1997Zero} noted the emergence of these stable/unstable shocks depending on the values of specific angular momentum, $\lambda$. More recent simulation studies have underscored the periodic luminosity variations associated with the Sgr A* cycle, attributing it to the oscillations of these shocks. For instance, the works of \citet{2019PASJ...71...49O,2022MNRAS.514.5074O} illustrated luminosity variability occurring over 5-10 day intervals. In addition to the mentioned supermassive black holes (such as Sgr A*), some black hole X-ray binaries (BHXRBs) have also reported low-frequency luminosity fluctuations\citep{2019A}. Among them, Type C low-frequency quasi-periodic oscillations are believed to be caused by fluctuations in accretion flow properties like mass accretion rates. Recently, \cite{2024MNRAS.tmp..220D} simulated accretion flows near a $10M_\odot$ black hole, revealing that shocks in the accretion flow can lead to the generation of flickering at several Hz. 

Building on the insights of previous research, this paper delves into the study of (in)stability of standing shocks in two-dimensional (2D) hydrodynamic low angular momentum accretion flows. Specifically, we revisit and extend the works by \cite{Dongsu1997Zero} with implementation of refined boundary conditions, computational domain utilizing advanced simulation software and modern mesh models. A comprehensive breakdown of these improvements can be found in Section \ref{sec:2}. Our approach is rooted in the conservation of specific energy (set to zero) and the conservation of specific angular momentum. These tenets guide our exploration of the instability inherent to standing shocks and help us gauge the variable nature of luminosity. To further cement the relevance of our study, we compare our findings with long-term observational data of the flares associated with Sgr A* and QPOs in some black hole X-ray binaries (BHXRBs).

The structure of this paper is as follows: Section \ref{sec:2} presents the methodologies driving our simulations, providing details related to the initial and boundary conditions. The results of our simulations and interpretations are presented in Section \ref{sec:3}. We culminate in Section \ref{sec:4} with a robust discussion, relating our findings within the broader landscape of existing knowledge and shedding light on their implications for ongoing research and conclusions in Section \ref{sec:5}.

\section{Numerical Methods}
\label{sec:2} 

\subsection{Basic Equations}
\label{sec:2.1} 

To conduct our simulations, we employ the PLUTO code \citep{2007jena.confR..96M}, which can handle evolution of high Mach number flows with strong discontinuities like shocks. The PLUTO code solves the time-dependent  conservation equations given by: $\frac{\partial\boldsymbol{U}}{\partial t} +\boldsymbol{\nabla} \cdot \boldsymbol{T_h} = \boldsymbol{\nabla} \cdot \boldsymbol{T_p}+\boldsymbol{S(U)}$, where $\boldsymbol{U}$ is a set of conservative quantities, $\boldsymbol{T_h}$ is flux tensor, $\boldsymbol{T_p}$ is diffusion flux tensor and $\boldsymbol{S(U)}$ is source terms. This framework is particularly useful to simulate supersonic flows across range of coordinate systems in multi-dimension. We consider an ideal
relativistic non-magnetized fluid with laboratory rest-mass density $\rho$, three-velocity $\boldsymbol {v}$, thermal (gas) pressure $P$, lorentz factor $\gamma$ and specific enthalpy $h$. The equations governing special relativistic hydrodynamics (RHD) are presented as \cite{1959flme.book.....L}:

\begin{equation}
\label{eq:1}
 \text{mass conservation equation}, \frac{\partial (\gamma \rho)}{\partial t}+\boldsymbol{\nabla} \cdot(\gamma \rho \boldsymbol{v})= 0,
\end{equation}
\begin{equation}
\text{momentum conservation equation}, \label{eq:2}\frac{\partial(\rho h \gamma^2  \boldsymbol{v})}{\partial t}+ \boldsymbol{\nabla} \cdot (\rho h \gamma^{2} \boldsymbol{v} \boldsymbol{v}  ) + {\nabla} P = -\rho \gamma^{4} \boldsymbol{v} ( \boldsymbol{v} \cdot {\nabla} \Phi) - \rho \gamma^{2} {\nabla} \Phi \hfill ,
\end{equation}
\begin{equation}
\text{energy conservation equation}, \label{eq:3}\frac{\partial E}{\partial t}+\boldsymbol{\nabla} \cdot\left(\rho h \gamma^2 \boldsymbol{v}\right)= -\rho \gamma^{4} v^{2} (\boldsymbol{v} \cdot {\nabla} \Phi) - \rho \gamma^{2} \boldsymbol{v} \cdot {\nabla}\Phi .
\end{equation}
Here, $E = \rho h \gamma^2-P$ and $\Phi$ represents the total energy and gravitational potential respectively.

\subsection{Energy Equation of Advective Flow}
\label{sec:2.2} 
For the advective flow, we consider the adiabatic equation of state $P = K\rho^\Gamma$. The self-gravity of the accreting gas is negligible in comparison to the gravitational influence of the central black hole \citep{1995MNRAS.272...80C}. Here, $\Gamma$ represents the adiabatic index, with a value of $4/3$, appropriate for a hot thermally relativistic single component fluid (ions) in inner advective flows around the black hole where temperature can reach up to even $10^{12}K$ while $K$ denotes gas constant. Additionally, the specific angular momentum, defined as $\lambda_0\sim R_\phi v_\phi$, can be considered conserved for flows with low angular momentum, especially in scenarios corresponding to the weak-viscosity limit \citep{Dongsu1997Zero}. Here,  $R_\phi$ is the projection of radial position on the equatorial plane and $v_\phi$ is the corresponding velocity.  Energy conservation gives us relation of specific energy of the flow or Bernoulli constant, $\varepsilon$ = $\frac{v_R^2}{2}$ + $\frac{c_{s}^2} {\Gamma -1}$ + $\frac{\lambda_0^2}{2 R^2}$ + $\Phi$ \citep{Dongsu1997Zero}. Here, $c_{s}$ being the sound speed.
Given the above mentioned considerations, the Bernoulli constant of the flow can be expressed as:

\begin{equation}
\label{eq:7}
\varepsilon = \frac{v_R^2 \sin ^2 \theta}{2}+\frac{v_R^2 \sin ^2 \theta}{\mathcal{M}^2(\Gamma -1)}+\frac{\lambda_0^2}{2 R^2}+\Phi.
\end{equation}

Here, $v_R$ represents the radial velocity, while $\mathcal{M}$ stands for the Mach number given by $\mathcal{M} = \frac{v_R}{c_{s}}$. Note that velocities are naturally expressed in units of the speed of light $c = 1$. The specific angular momentum is denoted by $\lambda_0$ and is expressed in units of $2GM/c$ and the radial distance $R$ in spherical co-ordinate is in units of Schwarzschild radius, $R_{g} = 2GM/c^{2}$. M being the mass of the
central compact object and G as the gravitational constant. To mimic the space-time around non-rotating black hole, we use the Pseudo-Newtonian potential, which serves as an approximation for the effective potential \citep{B1980Thick}:
\begin{equation}
\label{eq:8}
\Phi = -\frac{1}{2(R-1)}.
\end{equation}

Theoretical studies on transonic flows around black holes \cite[e.g.,][]{1989ApJ...347..365C,1990ttaf.book.....C,1993ApJ...417..671C} has provided detailed discussions on the conditions for shock formation with supersonic points concerning $\lambda_0$. Their findings indicate that, to satisfy the Rankine-Hugoniot condition, the specific angular momentum determine whether a physical supersonic point exists in the disc, represented by $\lambda_0 \geq \lambda_u\approx1.854$. Therefore, our work will revolve around this range of $\lambda_0$. Under these conditions, the flow's evolution is determined by four parameters set at the outer boundary of simulation computational domain: $v_R$, $\lambda_0$, $c_s$, and the half thickness of the accretion flow, $H$ which are determined by solving the equations of conservation laws.

\subsubsection{Vertical hydrostatic equilibrium assumption}
\label{sec:2.2.2}

\begin{figure}
\centering
\includegraphics[width=0.8\columnwidth]{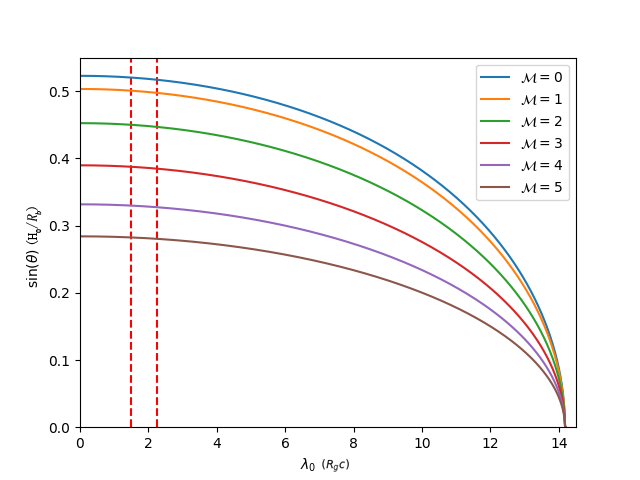}
\caption{\label{fig:2} This figure illustrates the relationship between the hydrostatic equilibrium height $ \frac{H}{R} $ and the specific angular momentum $ \lambda $ at a radius of $ 200R_g $ for different Mach number $ \mathcal{M} $ which are shown with different colours. In the extreme case where $ \lambda = 0 $ (corresponding to the Bondi flow), the equilibrium height approaches $ \sim 0.52 $. For each scenario, there is a rapid decline in the equilibrium height with increasing $ \lambda $, reaching zero around $ \lambda > 13 $ and becoming nearly negligible at $ \lambda \sim 14.2 $. The region delineated by the dashed red line corresponds to the specific angular momentum selection for our study, where the equilibrium height is approximately $ \sim 0.28 $.}
\end{figure}

The assumption of hydrostatic equilibrium has proven instrumental in the realm of advective flow studies. This fundamental concept was developed and notably advanced by Fukue and others \citep[e.g.,][]{1986PASJ...38..167F,2008bhad.book.....K}. Within this framework, the dimensionless gravitational acceleration is given by:\begin{equation}
\label{eq:10}
g_\theta= -\frac{\partial P}{R \rho \partial \theta} - \frac{\partial \Phi}{R \partial \theta} =-\frac{1}{2(R-1)^2}\frac{H}{R}sin\theta.
\end{equation}
Here, we take the gravitational acceleration ($g_\theta$) with the direction along $\theta$ (i.e., \textbf{$\theta_0$}) for convenience, and the term $\frac{H}{R}$ signifies the relative height. Under the one-zone approximation, which can be expressed as $ \frac{P}{H} \approx -\rho g_\theta /sin\theta $, the Eq. \ref{eq:10} can be rewritten as
\begin{equation}
\label{eq:10c}
\frac{P}{H} \approx \frac{\rho}{2(R-1)^2} \frac{H}{R}.
\end{equation}
Considering the dimensionless-adiabatic sound speed given by $ c_s = \sqrt{\frac{\Gamma P}{\rho}} $, the hydrostatic equilibrium in advective flows can be articulated as:
\begin{equation}
\label{eq:11}
\frac{H}{R}\approx\frac{c_s(R-1)}{\sqrt \Gamma}\sqrt{\frac{2}{R}}.
\end{equation}

In our investigation, we posit that the flow at the outer boundary adheres to hydrostatic equilibrium. Integrating this with the foundational equation \eqref{eq:7}, it emerges that the height ratio $ \frac{H}{R} $ at a fixed $R$ amplifies as the specific angular momentum, $ \lambda_0 $, diminishes. Within the paradigm of low-angular momentum advective flows, conventional computations position $ \frac{H}{R} $ at an approximate value of 28\%, as delineated in equation \eqref{eq:11}. Divergently, in the scenario of a extreme case $ \lambda_0 = 0 $ (synonymous with the Bondi flow) and the thin disk regime characterized by $ v_R \ll v_{\phi} $ (as seen in SSD), this proportion can escalate to around 52\% and down to 0\% respectively in radius 200 $R_g$ (see Fig. \ref{fig:2}). For all these estimations, the specific energy in equation \eqref{eq:7} is taken zero. 

\subsection{Initial Conditions}
\label{sec:2.3}
The disk height ratio $\frac{H}{R}$ of equation \eqref{eq:11} can also be obtained in another form for large $R (> > R_{g})$ as \cite{2019PASJ...71...49O}:
\begin{equation}
\label{eq:12}
\frac{H}{(R^2-H^2)^{3/4}} = 0.043(\frac{M}{10M_\odot})^{-1/2}(\frac{T}{10^{10}K})^{1/2}(\frac{1}{3\times10^6cm})^{1/2},
\end{equation}
where $M$ denotes the mass of the central compact object, for instance, a black hole. The term $ T $ is defined as $ T = \frac{c_s^2 \mu m_p}{\Gamma K_B} $, representing the temperature. Here, $ m_p $ stands for the proton mass, $ \mu $ signifies the average molecular weight (taking a value of 0.5 for hydrogen), and $ K_B $ is the Boltzmann constant. This shows the disk-height ratio is small near the inner edge of the flow but as large as $\sim 0.37$ at $R\sim 200 R_{g}$ for $T \sim 10^{10}K$. Furthermore, there exists an inverse relationship between temperature $ T $ and the Mach number $ \mathcal{M} $; a larger $ \mathcal{M} $ corresponds to a reduced $ T $. To ensure the detection of at least one shock with appreciable intensity in our simulations, we opted for $ \mathcal{M} = 5 $. This choice positions the shock location in the range $ R_s \sim 10 - 100 R_g $ for every scenario considered. 
Drawing from the standard analysis of transonic flow by Chakrabarti and collaborators, and based on previous works, we adopt specific angular momentum values ranging from 1.5 to 2.25, incremented by 0.05. Subsequently, we use the pseudo-Newtonian potential (given by Eq. \ref{eq:8}) into the energy equation (Eq. \ref{eq:7}). The governing equation governing our simulations is thus:
\begin{equation}
\label{eq:13}
\frac{v_R^2 \sin ^2 \theta}{2}+\frac{v_R^2 \sin ^2 \theta}{\mathcal{M}^2(\Gamma-1)}+\frac{\lambda_0^2}{2 R^2}-\frac{R_g}{2\left(R-R_g\right)} = 0.
\end{equation}

Once the dimensions of the simulation box are specified, we can derive the velocity distribution at the outer boundary. Opting for the smallest value in this distribution, specifically $ \theta = \frac{\pi}{2} $, enables us to compute the initial temperature. By assuming that the initial environment possesses a notably low dimensionless density, represented as $ \rho_\infty = 0.01 $, we can subsequently ascertain another basic parameter: the gas pressure.

\begin{figure}
\centering
\includegraphics[width=0.8\columnwidth]{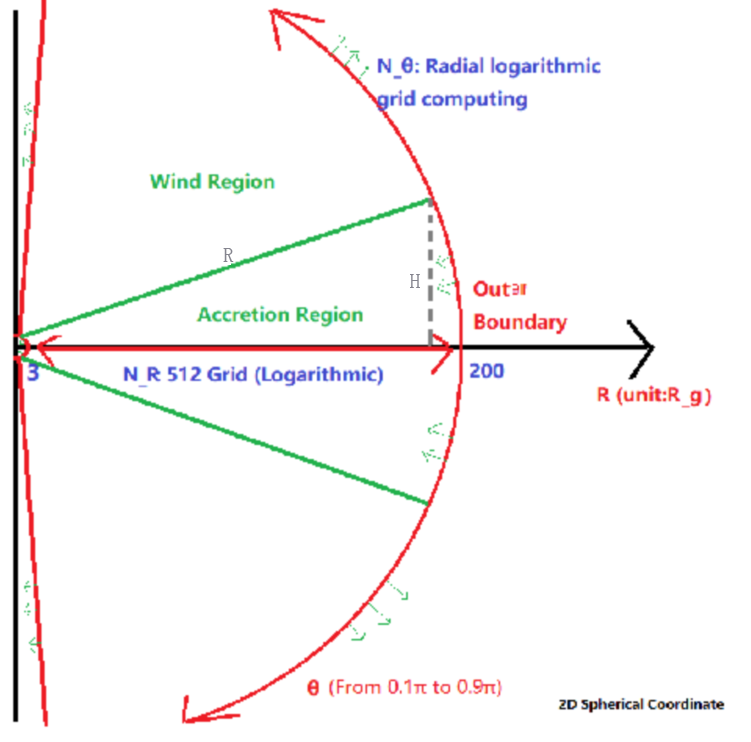}
\caption{\label{fig:3} The boundary conditions for our simulations are represented in the following manner: The red and blue lines illustrate the boundaries of the simulation box and their associated grid numbers. Concurrently, the green line designates the division between the wind and accretion regions, with the green arrow signifying the flow direction.}
\end{figure}
 
\subsection{Boundary Conditions}
\label{sec:2.4}
Utilizing the 1.5D solutions with $ \mathcal{M} = 5 $ in conjunction with equation (\ref{eq:13}), we set the outer boundary at $ R_{\text{out}} = 200R_g $. To circumvent any void regions, particularly within the innermost stable radius, we establish the inner boundary $ R_{\text{in}} = 3 R_g$. For the $ \theta $ direction, to avoid computationally frequently arising negative density or pressure regions, we restrict $ \theta $ to range from $ 0.05\pi $ to $ 0.95\pi $. Spatially, we employ 512 logarithmic grids in the ``$R$'' direction. The use of non-uniform grid like logarithmic one in the radial direction allows us to keep the mesh aspect ratio same throughout the computational domain. Besides, following condition keeps the aspect ratio around unity i.e., almost squared cells throughout the domain \citep{2007jena.confR..96M}:

\begin{equation}
\label{eq:14}
\Delta R_1\approx R_1 \Delta \theta = R_1 \frac{\theta_\text{end}-\theta_\text{beg}}{N_\theta},
\end{equation}
where $\Delta R_1$ is the first active mesh size in the radial direction and $\theta_\text{end}$ and $\theta_\text{beg}$ are the first and last zone in the ``$\theta$'' direction respectively. $N_\theta$ is the number of computational zones in the ``$\theta$'' direction. We can determine either the number of points in the radial direction or the endpoint as follows,
\begin{equation}
\label{eq:15}
\log_{10} \frac{R_\text{out}}{R_{\text{in}}} = N_r \log _{10} \frac{2+\Delta \theta}{2-\Delta \theta}.
\end{equation}
where $\theta_\text{end}-\theta_\text{beg} = 0.9\pi$ in our simulation and the subscript "1" represents the first cell. $N_r$ is the number of computational zones in the radial direction ``$R$'' .

For our inner boundary, we employ an outflow (without inflow, i.e., we make sure that all the infalling matter achieves  either zero or negative value but not positive value in radial direction that means that no inflow occurs from ghost cells to physical cells) condition while preserving angular momentum, effectively replicating the role of the region within the innermost stable radius. In the wind region, we enforce an outflow condition. Meanwhile, for the accretion region, we impose an inflow (with no outflow) condition, using velocities determined by the governing equation (\ref{eq:13}) and setting a dimensionless density $ \rho_{\text{in}} = 1 $. The demarcation between the wind and accretion regions is ascertained using the hydrostatic equilibrium equation (\ref{eq:11}), which is represented by the green line in Fig. \ref{fig:3}.  In our simulation set up, the injection parameters for accretion flow is provided in the region with the thickness above and below the equation plane given by the equation (\ref{eq:11}) and in the rest of the region, it is imposed that there is only outflow and no inflow of matter. 

\section{Numerical Results}
\label{sec:3}
 \begin{table}
     \centering
     \begin{tabular}{cccccccccc}
    \hline
    $\lambda_0$ & $c_{s,\infty}$ & $T_\text{min}$ & $P_\infty$ & $T_\text{end}$ \\
    ($R_gc$) & ($c$) & ($K$) & ($\rho_u c^2$) & ($t_0$) \\
    \hline
    1.50 & $1.272\times 10^{-2}$ & $6.56\times 10^8$ & $1.213\times 10^{-6}$ & $1 \times 10^5$ \\
    1.60 & $1.271\times 10^{-2}$ & $6.55\times 10^8$ & $1.211\times 10^{-6}$ & $1 \times 10^5$ \\
    1.65 & $1.270\times 10^{-2}$ & $6.54\times 10^8$ & $1.210\times 10^{-6}$ & $1 \times 10^5$ \\
    1.70 & $1.270\times 10^{-2}$ & $6.53\times 10^8$ & $1.209\times 10^{-6}$ & $1 \times 10^5$ \\
    1.75 & $1.269\times 10^{-2}$ & $6.53\times 10^8$ & $1.208\times 10^{-6}$ & $2 \times 10^5$ \\
    1.80 & $1.269\times 10^{-2}$ & $6.52\times 10^8$ & $1.207\times 10^{-6}$ & $5 \times 10^5$ \\
    1.85 & $1.268\times 10^{-2}$ & $6.52\times 10^8$ & $1.206\times 10^{-6}$ & $7 \times 10^5$ \\
    1.90 & $1.267\times 10^{-2}$ & $6.51\times 10^8$ & $1.205\times 10^{-6}$ & $2 \times 10^5$ \\
    1.95 & $1.267\times 10^{-2}$ & $6.51\times 10^8$ & $1.205\times 10^{-6}$ & $5 \times 10^5$ \\
    2.00 & $1.266\times 10^{-2}$ & $6.50\times 10^8$ & $1.202\times 10^{-6}$ & $5 \times 10^5$ \\
    2.25 & $1.263\times 10^{-2}$ & $6.46\times 10^8$ & $1.196\times 10^{-6}$ & $1 \times 10^5$ \\
    \hline
  \end{tabular}
  \caption{ \label{tb:1} 
  A set of initial parameters: specific angular momentum $\lambda_0$, initial sound speed $c_{s,\infty}$, minimal temperature $T_\text{min}$, initial gas pressure $P_\infty$. All cases with initial dimensionless density $\rho_\infty = 0.01$, inflow dimensionless density $\rho_\text{inj} = 1.0$, mass of black hole $M = 10 M_{\odot}$ and the geometric height of the inflow $H_0/R_b\sim0.28$ with $\mathcal{M_\infty} = 5$ (see Fig.\ref{fig:2}), where $R_{b}$ represents the outer boundary in the radial direction. The simulations time ($t_\text{end}$) without shocks will run to $10^5$ (units of $R_g/c$, $t_0$ thereafter), and $t_\text{end}$ is flexible with shocks according to the disk situations.}

 \end{table}

\begin{figure}
\centering
\includegraphics[width=\columnwidth]{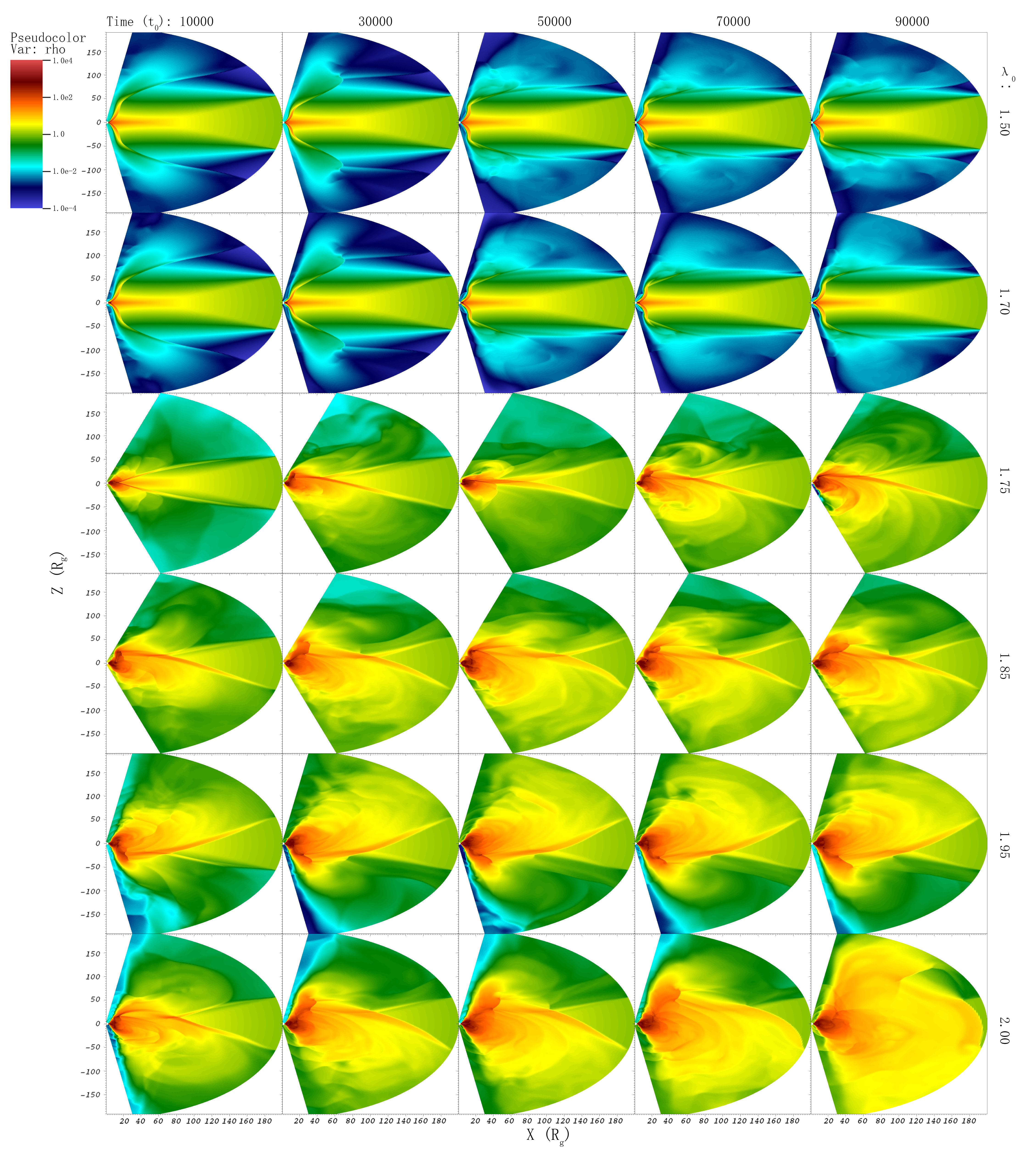}
\caption{\label{fig:10} Time evolution of 2D density profiles for various angular momentum, $\lambda_0$, namely 1.50, 1.70, 1.75, 1.85, 1.95, and 2.00 where the evolutionary times are given at  10000, 30000, 50000, 70000, and 90000$t_0$. For $\lambda_0 = 1.75$ and 1.85, we adopted a bigger cut in the funnel region to circumvent numerical errors associated with $v > c$. Generally, a more pronounced disk wind is particularly noticeable for $\lambda_0 = 1.75$ where the envelope gets disrupted by this wind.}
\end{figure}

Based on the above initial and boundary conditions, we conduct 2D simulations of the low angular momentum advective flow. The initial values, such as velocity and pressure, are derived from the governing equation (\ref{eq:13}) and the hydrostatic equilibrium equation (\ref{eq:11}). The model parameters and the initial variables at the outer boundary are detailed in table \ref{tb:1}. The mass fluxes of the accretion and wind, represented by $ \dot{M} = \iint \rho v \cdot d \boldsymbol{S} $, across the boundaries, as illustrated in Fig. \ref{fig:3}. Fig. \ref{fig:10} presents the evolution of the density profile across different specific angular momentum values $ \lambda_0 $, namely 1.50, 1.70, 1.75, 1.85, 1.95 and 2.00. 

\begin{figure}
\centering
\includegraphics[width=0.8\columnwidth]{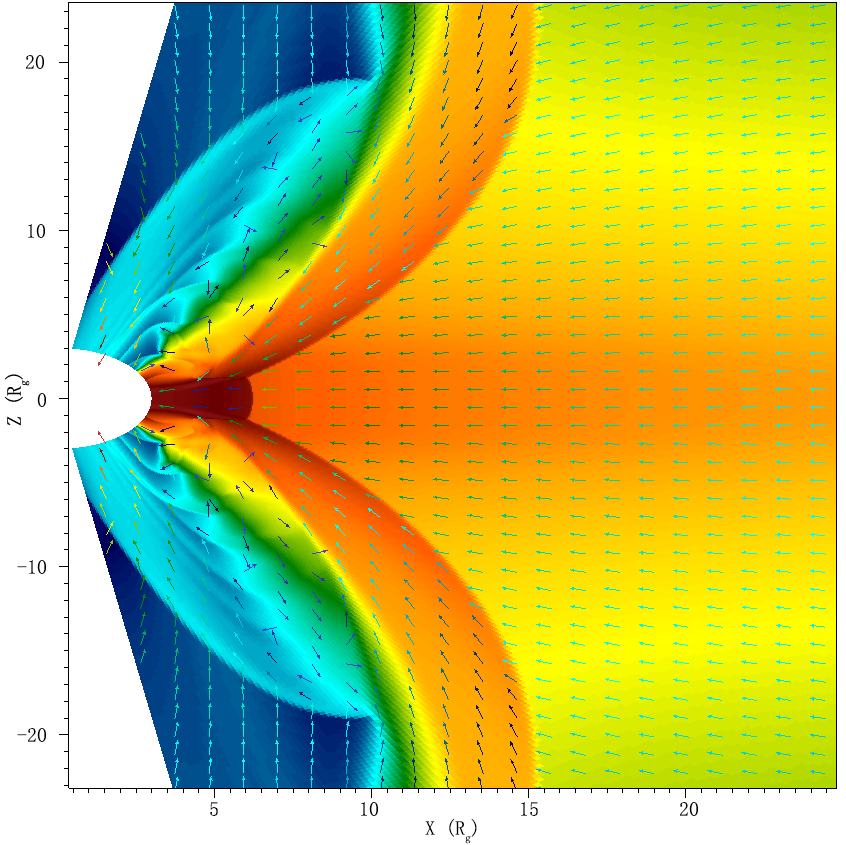}
\caption{\label{fig:14} Density profile and velocity distribution (indicated by arrows) for $\lambda_0 = 1.70$ at time $70000t_0$. The X and Z axes are in units of ($R_g$) respectively. Intermittent shocks are produced due to the collision of inflows and outflows in the funnel region.}
\end{figure}

This equilibrium height closely approximates 0.28, with further details provided in Fig. \ref{fig:2} and table \ref{tb:1}. 
For lower angular momenta, such as $\lambda_0 \leq 1.70$, the accretion process remains highly stable, even with a standing shock and the oblique shocks near the inner boundary (see Fig. \ref{fig:14}). As the specific angular momentum increases, leading to the emergence of shocks, the symmetry is initially compromised, followed by the formation of vortices within the accretion disk. This results in a more substantial accumulation of matter. Interestingly, accretion disks with very large angular momenta, despite having more robust disk winds, manifest a relatively stable state when compared to their lower angular momentum counterparts. Additionally, for $\lambda_0 \geq 1.75$, the robust disk wind disrupts the envelope of the accretion flow , ultimately forming a unidirectional vortex structure. Furthermore, when $\lambda_0 \geq 2.00$, the vortex structure becomes dynamically unstable, and the powerful disk wind causes the size of the vortex region to continuously increase over time.

\begin{figure}[H]
\includegraphics[width=0.9\columnwidth]{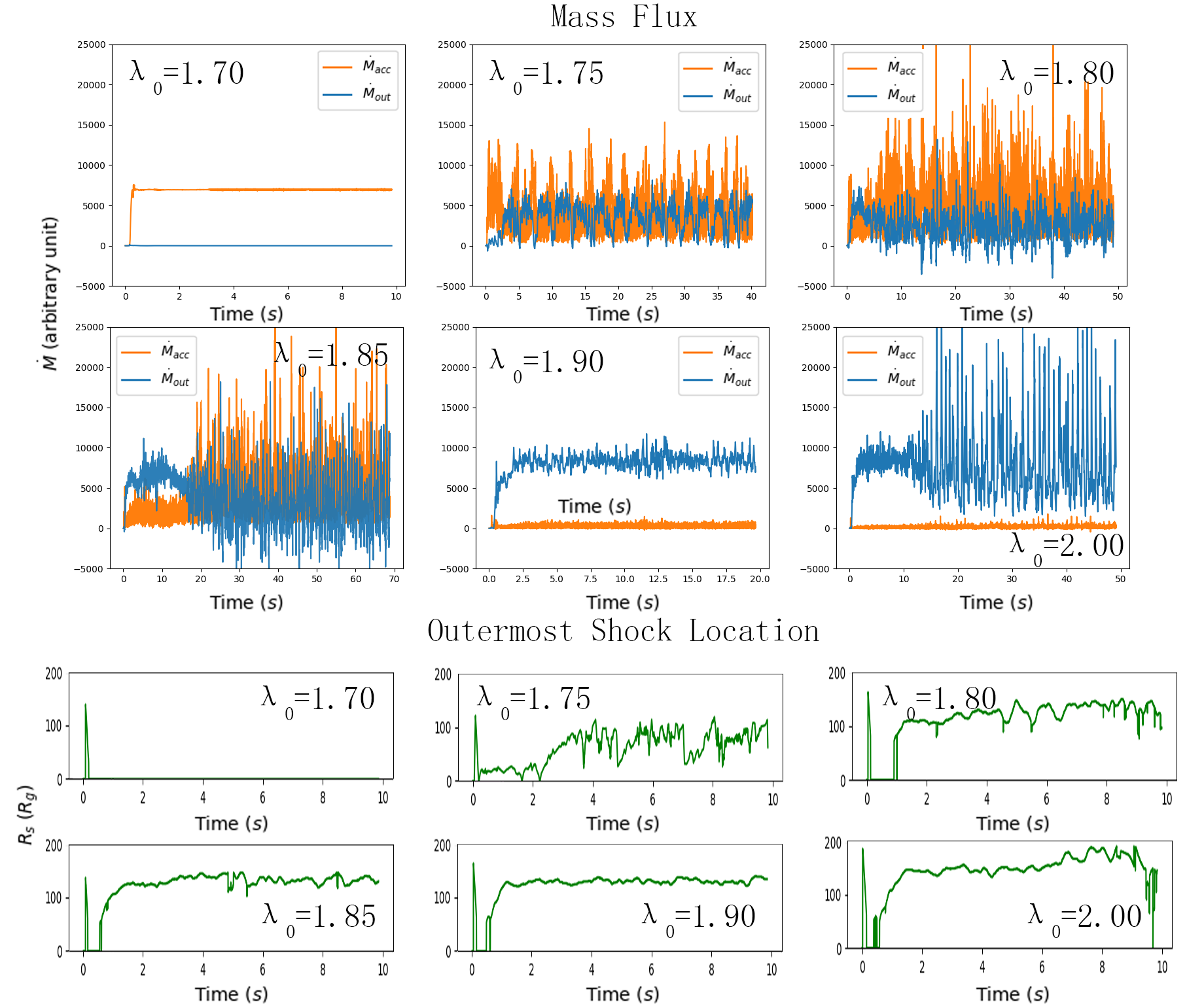}
\caption{\label{fig:11} Top planes: Time evolution of the accretion rate ($\dot{M}_\text{acc}$, orange line with arbitrary unit), outflow rate ($\dot{M}_\text{out}$, blue line with arbitrary unit) with specific angular momentum 1.70, 1.75, 1.80, 1.85, 1.90 and 2.00 across whole simulations. Both mass flux derived from derived from $ \dot{M} = \iint \rho v \cdot d \boldsymbol{S} $. Bottom planes: Outermost shock location on the equatorial plane ($R_s$, green line) within $10^5 t_0$ (i.e., about ten seconds) after beginning of the simulations for above cases.  A detection threshold of 10\% has been employed in measuring $R_s$.  With increasing specific angular momentum, the dynamics of accretion and disk wind in the system evolve. Notably, for $\lambda_0 \geq 1.85$, the $\dot{M}_\text{out}$ substantially exceeds the $\dot{M}_\text{acc}$, making it challenging for material to descend into the black hole. For very low values of $\lambda_0$, as the shock position is very close to the inner boundary, we employed a higher signal filtering strength to avoid distortion effects. For samples exhibiting shocks (i.e., $\lambda_0\geq1.75$), oscillations in both the $\dot{M}_\text{acc}$ and $\dot{M}_\text{out}$ are primarily attributed to these shocks.  As $\lambda_0$ amplifies, the shocks' location approaches the outer boundary, exhibiting an inverse trend with the accretion rate, characterized by a discernible time lag. While oscillating shocks are observed in these samples, the shock locations for $\lambda_0 \geq 2.00$ display outward oscillations over time, indicating their inability to sustain a stable quasi-periodic oscillation state. In the late evolution at $\lambda_0 = 2.00$, the sudden decrease in the evolution plot of $R_s$ is due to distortion caused by the outermost shock extending beyond the outer detecting boundary. However, it should be noted that for $\lambda_0 = 1.70$, the observed shock is primarily associated with the collision of inflow and outflow near the inner boundary, rather than within the accretion region itself (see Fig. \ref{fig:14}).}
\end{figure}

\begin{landscape}
    \begin{figure}
    \centering
    \includegraphics[width=0.8\columnwidth]{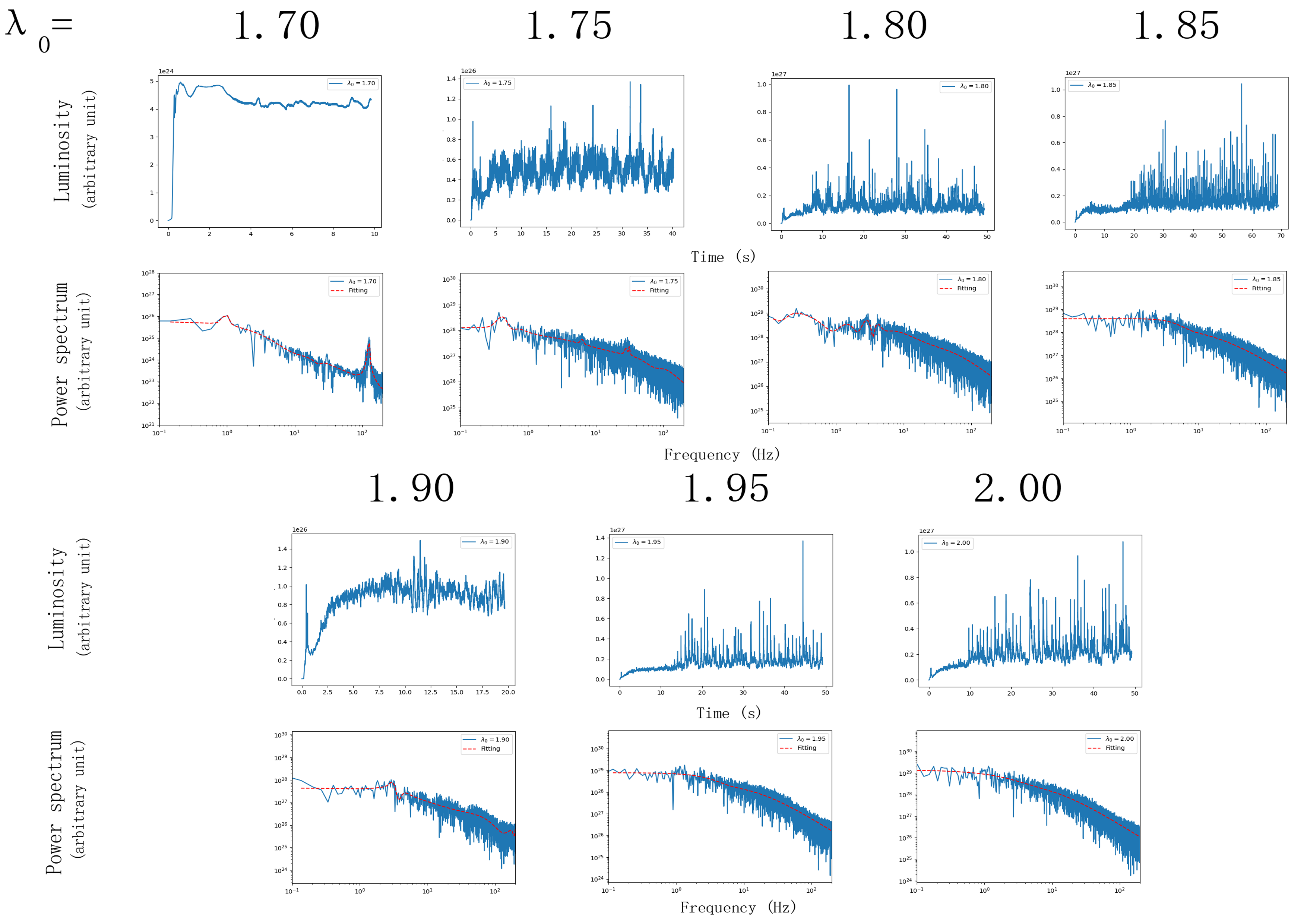}
    \caption{\label{fig:13} The panels display the light curve derived from Bremsstrahlung (see equation \ref{eq:16}), and the power spectrum obtained by fast Fourier transform (FFT, blue solid line) and its nonlinear multiple model fitting (red dash line) for specific angular momenta $\lambda_0$ of 1.70, 1.75, 1.80, 1.85, 1.90, and 2.00. Notably, luminosity oscillations are discerned around frequencies of 120 and 1 Hz for $\lambda_0$= 1.70, 0.4, 5.0 and 30 Hz for $\lambda_0$= 1.75, and 0.25 and 3.0 Hz for $\lambda_0$= 1.80. Generally, the luminosity levels span between $10^{30}$ and $10^{33}$ erg/s with inflow density $10^{12}m_p$ and $g_{ff}$ of unity value for simplicity, aligning well with observational findings of a series of low-mass BHXRBs \citep{2019A}, and $10^{33}$ to $10^{37}$ erg/s for typical gas densities around BHXRBs in the range of $\sim$ $10^{10}$ to $10^{11} \text{g}\cdot \text{cm}^{-3}$ \citep[e.g.,][]{2021MNRAS.507..330S,2021ApJ...920..120Y}.}
    \end{figure}
\end{landscape}

For an insight into the accretion rate and the magnitude of the outflow, mainly due to the disk wind across the specified cases, readers can refer to Fig. \ref{fig:11}. Initial stages of the simulation witness a sharp escalation in the mass flux, attributable to the inflow first approaching the inner boundary. The dynamic stabilization time will increase as $\lambda_0$ increases (see Fig. \ref{fig:13}). Intriguingly, as $ \lambda_0 $ amplifies, the accretion rate demonstrates a declining trajectory, contrasting the substantial ascent in outflow. Particularly, for $ \lambda_0 \geq 1.75 $, oscillations in varied amplitudes mark both the accretion and outflow rates, signifying the presence of shocks. Shocks are detected for $\lambda_0 \geq 1.75$, with varying intensity in the range $1.75 \leq \lambda_0 \leq 1.80$. Cases with $\lambda_0 \geq 1.85$ show a tendency towards stable behaviour, while for $\lambda_0 \geq 2.00$, although oscillatory behavior is observed, it tends to move outward, even extending beyond the detection region. 

In the simulations, we observed a different behavior of the shock positions and mass accretion rates. Specifically, after a lag time during declining accretion rates, the shock positions move away from the black hole. This phenomenon is also supported by other numerical simulations \citep{2019PASJ...71...49O,2022MNRAS.514.5074O}. Moreover, these studies delve into the time intervals between changes in accretion rates and shock positions, attributing them to the propagation of acoustic waves. One possible explanation is that mass accretion rates are calculated near the inner boundary close to the black hole, while shock positions are typically tens of $R_g$ away from the inner boundary. This suggests that acoustic information does not synchronize between these regions. Observations of Sgr A* have also reported several hours of time lag between X-ray emissions and emissions from other regions in the accretion disk, such as submillimeter and infrared bands \citep{Yusef-Zadeh_2006, 2008ApJ...682..361Y, Yusef-Zadeh_2009}.

With smaller angular momentum values (specifically, $ \lambda_0 \leq 1.7 $), the accretion rate $\dot{M}_\text{acc}$ is predominantly higher. While there are discernible weak oscillations, given the absence of shocks in these cases, such oscillations are surmised to arise from the counter-effects of disk wind. For cases of moderate angular momentum (i.e., $ \lambda_0 = 1.75$ and $1.80$), the amplitude of the $\dot{M}_\text{out}$ mirrors that of the $\dot{M}_\text{acc}$. Nonetheless, considering the consistent oscillation of the $\dot{M}_\text{out}$ around zero, the dominant narrative remains that of material being predominantly channeled into the black hole. In contrast, for higher angular momentum values (specifically, $ \lambda_0 \geq 1.85 $),  $\dot{M}_\text{out} > \dot{M}_\text{acc}$. This delineates a scenario where the bulk of the material do not fall into the black hole, and is predominantly propelled outward as disk wind, in particular $\lambda_0 \geq 1.90$.

\begin{figure}
\centering
\includegraphics[width=0.8\columnwidth]{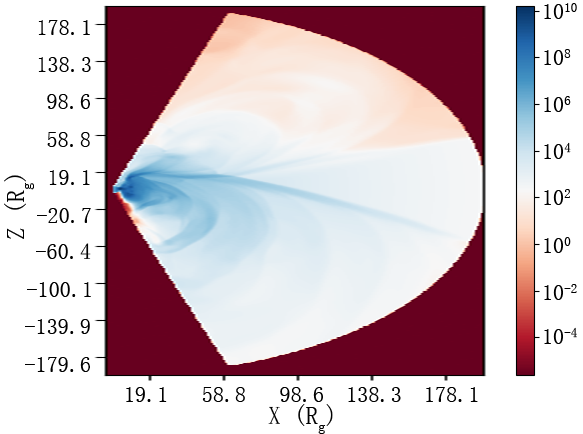}
\caption{\label{fig:8} The distribution diagram of Bremsstrahlung
as depicted in equation \ref{eq:16}, at a time of $ 90000t_0 $ for $ \lambda_0 = 1.75 $ under showcases characteristics reminiscent of Fig. \ref{fig:10}  (last column of the third row), with notably similar outlines. The unit in the diagram is arbitrary, where we considered frequency average of the velocity averaged Gaunt factor $g_{ff}$ as unity for simplicity. Particularly in the dense regions, which are proximate to the inner boundary along the equatorial plane (corresponding to the reddish area), are significant contribution to the Bremsstrahlung radiation.} 
\end{figure}

We consider here Bremsstrahlung emission, as an important radiation mechanism if the gas is optically thin to the Bremsstrahlung emission. The luminosity defined as \citep{1979Lightman}:
\begin{equation}
\label{eq:16}
L_\text{br} = g_{ff}
\int 1.4 \times 10^{-27}\left(\frac{\rho}{m_p}\right)^2 T^{\frac{1}{2}} d V,
\end{equation}
where $ \rho $ is the gas density, $ T $ is the temperature and $g_{ff}$ is the frequency average of the velocity averaged Gaunt factor with numerical value in the range of 1.1 to 1.5. We plotted the bremsstrahlung distribution diagram in Fig. \ref{fig:8} at 90000$t_0$. It is evident that regions with higher density contribute significantly to the Bremsstrahlung emission rate per unit volume, even showcasing a similar profile. As such, the shock-dominated evolution model can directly influence the distribution of Bremsstrahlung radiation.

We examine the light curves derived from equation \ref{eq:16} for various values of specific angular momentum, namely $\lambda_0 = 1.70, 1.75, 1.80, 1.85, 1.90, 1.95$, and $2.00$, as depicted in Figure \ref{fig:13}. Since low-frequency oscillation peaks are more susceptible to noise, one oscillation peak may be decomposed into multiple oscillation peaks in the same area. In order to restore the original state of the oscillation peak as much as possible, we considered three sources of radiation in the fitting: Lorentz, power law and Gaussian.  The Lorentz spectrum is usually related to the radiation process under the relativistic effect; power-law spectrum is usually related to the non-thermal radiation process in the accretion disk, mainly related to the magnetic field; and Gaussian spectrum is mainly related to certain spectral lines. As there 

\begin{figure}[H]
\centering
\includegraphics[width=0.8\columnwidth]{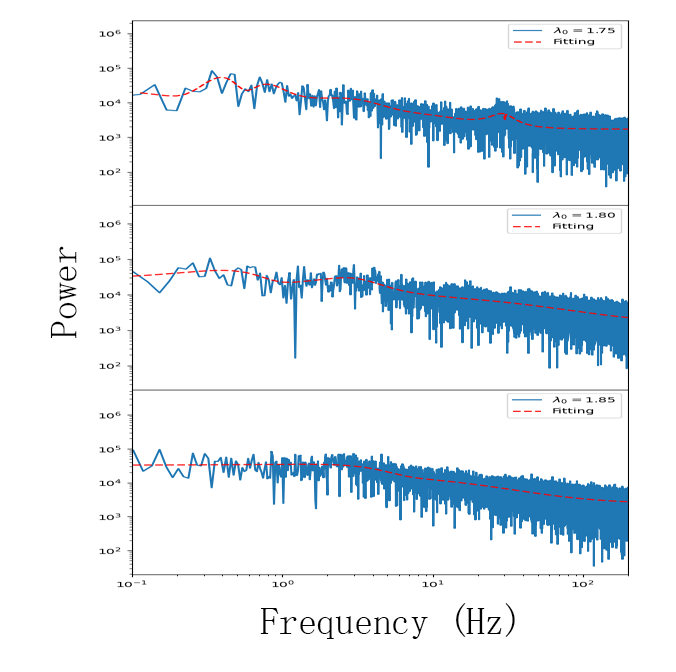}
\caption{\label{fig:15} The oscillating spectrum (blue solid line) of the outermost shock location and its nonlinear multiple models fitting (red dash line). The x-axis represents frequency (in Hz), and the y-axis represents relative amplitude. The four sub-figures correspond to values of $\lambda_0$ with 1.75 (top), 1.80 (middle), 1.85 (bottom).}
\end{figure}

{
\noindent
are not magnetic field in the simulations, the spectrum is mainly contributed by the Lorentz and the Gaussian spectrum, which are the main evaluation items of the fitting results. In addition, for all fitting results ploted in Fig. \ref{fig:13}, we found $\chi^2$ to be close to 1, especially in the low frequency range (except the case with $\lambda_0 = 1.70$ has a strong oscillation peaks in the high frequency band) which corresponds to the frequency range of LFQPOs. Pronounced luminosity oscillations around few Hz are discernible for $\lambda_0 = 1.70, 1.75$, and $1.80$. It's essential to realize that the oscillations at $\lambda_0 = 1.70$ are not directly a consequence of unstable shocks within the accretion region, there are even standing shock and oblique shock near inner boundary (see Fig. \ref{fig:14}). Rather, they are primarily caused by intermittent shocks generated by the rapid collision of outflowing disk winds and the infalling flow within the funnel region and outside the equipotential surface. For $\lambda_0\geq1.85$, the high angular momentum causes matter to continuously accumulate within the accretion flow, leading to relatively stable shock (refer to Fig. \ref{fig:11}), which may be the reason why the QPOs cannot be detected in these cases.
}

In an effort to investigate the relationship between luminosity and shock position, we examined the oscillation profile of the shock position for $\lambda_0$ values of 1.75, 1.80 and 1.85 (as depicted in Figure \ref{fig:15}). Notably, for $\lambda_0 \leq 1.80 $ exhibit frequencies and relative amplitudes that align with the findings shown in Figure \ref{fig:13}. Intriguingly, for $\lambda_0 = 1.85$ while oscillations of the shock position were detected, no distinct peak was observed in the power spectrum (Figure \ref{fig:13}). This observation might explain why pronounced oscillations are evident in $R_s$. Interestingly, \cite{Dhang2016Spherical,2018Magnetized} simulated the thermal accretion flow near neutron stars and reported similar oscillatory luminosity behavior. However, their analysis was limited to the shock range, revealing a consistent, high-frequency luminosity oscillation within that domain, hinting at high-frequency QPOs ranging from hundreds to a thousand Hz. This observation might serve as an indicator for the potential discovery of high-frequency QPOs in our system. In conclusion, our model provides a basis for understanding the LFQPO contribution from oscillatory shocks in the 0.1-30 Hz range for BHXRBs.

\begin{figure}
\centering
\includegraphics[width=0.8\columnwidth]{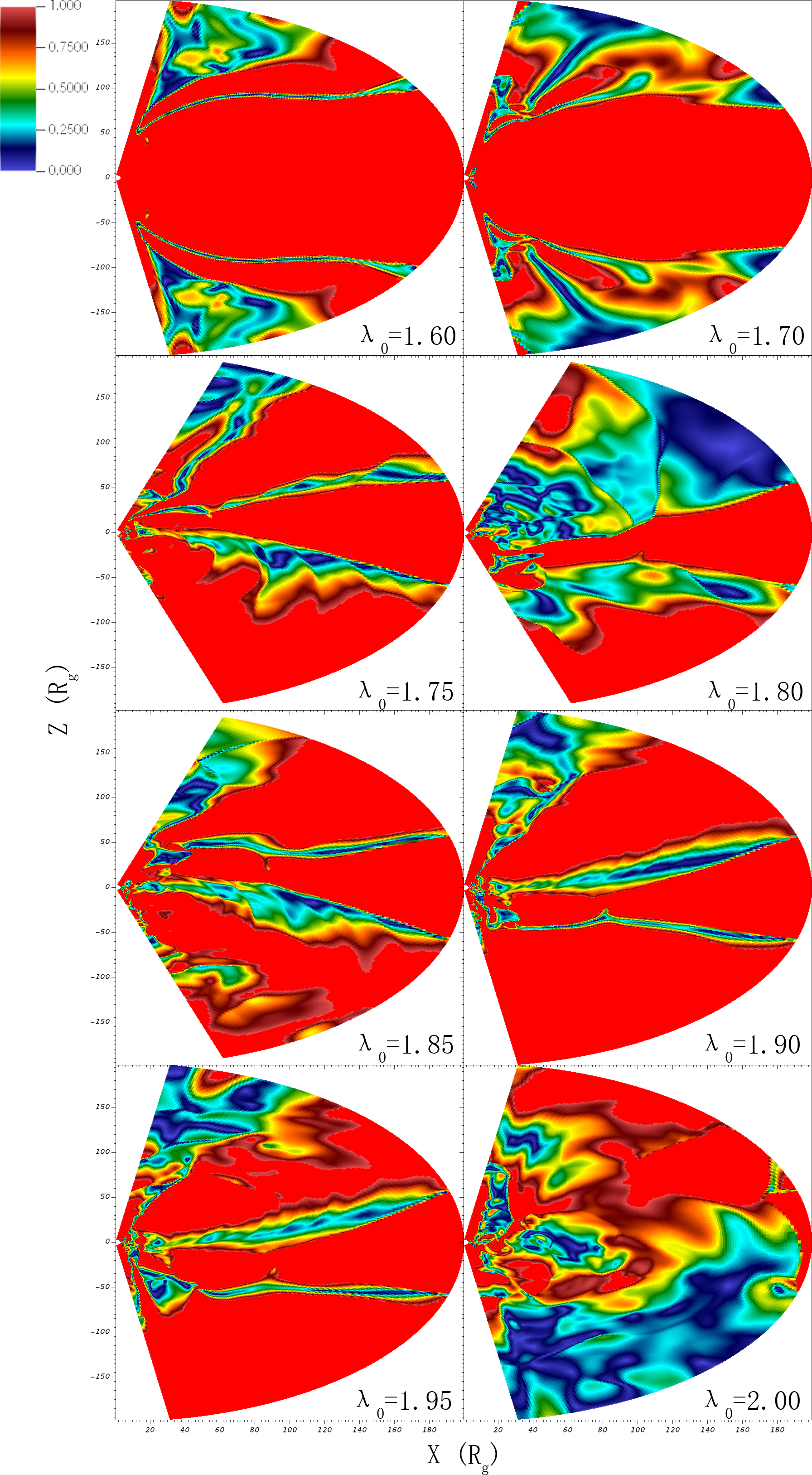}
\caption{\label{fig:21} The distribution of Mach numbers at $t=90000t_0$ for different values of $\lambda_0 =$1.60, 1.70, 1.75, 1.80, 1.85, 1.90, 1.95, and 2.00. To distinguish between subsonic and supersonic flows, all supersonic flows are shown in red color (regardless of magnitude). Arrows in the figure indicate the direction of accretion flow.}
\end{figure}

 \section{Discussion}
 \label{sec:4}
 In Section \ref{sec:3}, we presented the simulation results for the parameters listed in Table 1. In this section, for a more in-depth investigation into the stability of shocks and accretion flow, we delve into further discussions. In figure \ref{fig:13}, it can be seen that the luminosity of some cases exhibits burst behavior at certain moments, and in the long run, this burst seems to have some quasi-periodicity. Since there is no outburst phenomenon when shocks do not appear, we believe that this may be related to the activity mechanism of the accretion flow when shocks exist. Among them, figure \ref{fig:13} also shows that there is a weak shock, but there is no rapid variation of luminosity in $\lambda_0 = 1.70$. In contrast, $\lambda_0\geq1.75$ has different degrees of variations, and has more obvious shock and significant convection. Our results in section \ref{sec:4.1} shows that multiple shocks can exist in the accretion flow and that convection may cause the merger of two shocks and a short-term rise in the accretion flow behaviour. In section \ref{sec:4.2}, we further discussed the properties of accretion flow.
 
\subsection{Transonic Points and Multiple Shocks}
 \label{sec:4.1}
 
According to the Rankine-Hugoniot conditions, the pre-shock accretion flow must be supersonic (i.e., $\mathcal{M}>1$), and the post-shock flow must be subsonic. Therefore, a reasonable transonic point with multiple sonic points must exist for a shock to be present. In other words, the position of the shock should coincide with the location predicted by theory of the transonic flows. The theoretical model for transonic accretion flow suggests that the Rankine-Hugoniot condition for shock formation is satisfied for $\lambda_0>1.854$ (i.e., a physical sonicpoint). To discuss why shocks occur at angular momenta less than $\lambda_u$, we plotted the Mach number distribution (Fig. \ref{fig:21}). For very low $\lambda_0$, the accretion flow almost freely falls into the black hole. Everything undergoes a transition at $\lambda_0 \geq 1.75$, at which point multiple sonic points begin to appear (manifested in the change from red to blue regions in the radial direction in the figure). As the study of transonic flows is based on 1D models, and as shown in Fig. \ref{fig:21}, the distribution of sonic points in the plane is not confined to a specific radial direction. Therefore, we anticipate more occurrences of physically relevant sonic points in higher-dimensional scenarios. Although Fig. \ref{fig:21} does not reflect the relative strength and stability of shocks, when combined with Fig. \ref{fig:13}, shocks appear more stable for $\lambda_0 \geq \lambda_u$.

\begin{figure}
\centering
\includegraphics[width=0.8\columnwidth]{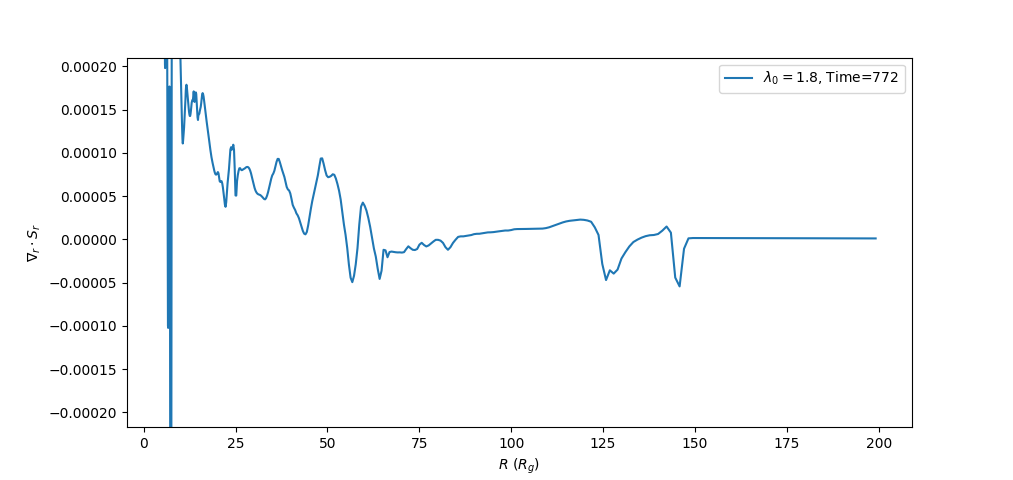}
\includegraphics[width=0.8\columnwidth]{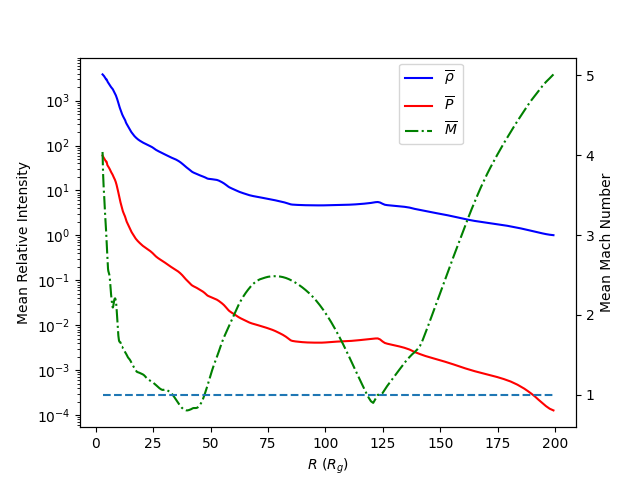}
\caption{\label{fig:17} Radial entropy divergence distribution ($\nabla_R \cdot S_R$, top) and average mach number ($\overline{\mathcal{M}}$, green dash-dot line) - density ($\overline{\rho}$, blue solid line) - pressure ($\overline{P}$, red solid line) diagram (bottom) for $\lambda_0 = 1.80$ at time $77200t_0$. The average value is obtained in the $\theta$ direction. Each negative peak of $\nabla_R \cdot S_R$ with $\mathcal{M}<1$ (see the blue dash line in the bottom plane), and rapidly increasing of $\overline{\rho}$ and $\overline{P}$ indicate a potential location for shock emergence. In this sample, the shocks are potentially located in the radial direction at $R \sim 50$ and $125R_g$.}
\end{figure}

In case of theoretical solutions, there is only one stable, physical shock along with multiple sonic points even when $\lambda_0 > 1.854$ \citep{1990ttaf.book.....C}. According to the physical characteristics of shock, the flow parameters will undergo a jump between the pre-shock flow and post-shock flow. Mathematically, this will manifest as derivatives resembling Dirac delta functions, making the divergence distribution a good criterion for determining the existence of shocks. Fig. \ref{fig:17} describes the divergence distribution in the radial direction ($\nabla_R \cdot S_R$) and mean mach number ($\overline{\mathcal{M}}$) - density ($\overline{\rho}$) - pressure ($\overline{P}$) (the quantities being averaged over $\theta$ direction) diagram at $77200t_0$ for $\lambda_0 = 1.80$. The shocks are detected around 50 and 125 $R_g$. Furthermore, on the bottom plane in Fig. \ref{fig:17}, although $\mathcal{M}<1$ only appears once at $R\sim50R_g$, but due to the average count and the range of $R$ of $\mathcal{M}<1$ as long as $20R_g$, combined with the $\nabla_R \cdot S_R$ graph, it can be potentially predicted that there is more than one shocks from $30$ to $50R_g$.

\begin{figure}
\centering
\includegraphics[width=0.8\columnwidth]{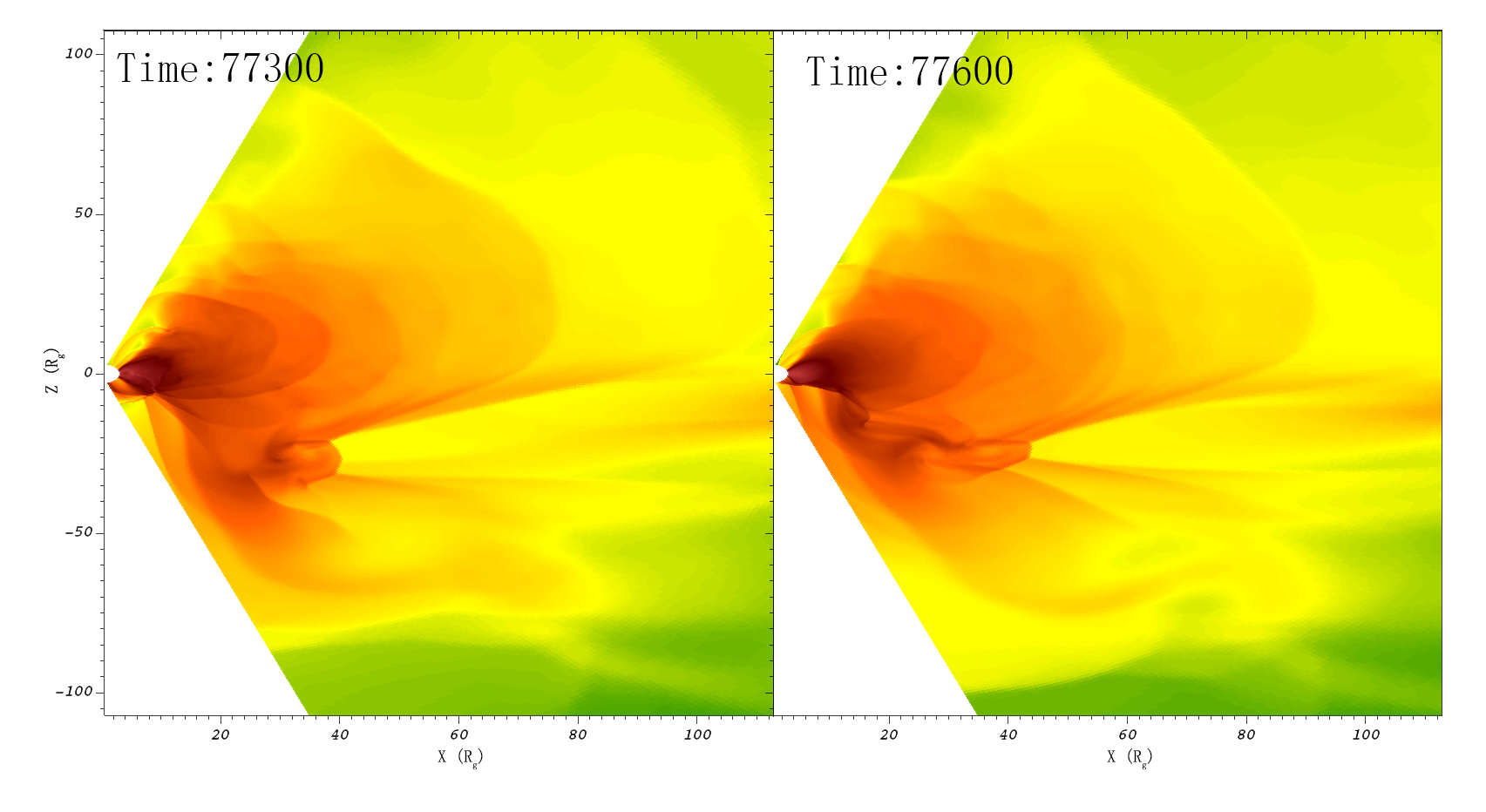}
\caption{\label{fig:12} The evolution of the density profile for $\lambda_0 = 1.80$ at 77300$t_0$ (left) and 77600$t_0$ (right).
}
\end{figure}

Simulations and discussions by \cite{2022MNRAS.514.5074O} reported an inner shock called the expanding shock, which contributes to the outflow process and affects the outer shock with opposite evolutionary behavior.  The behaviour of multiple shock waves is closely related to the local properties of the accretion flow. Fig. \ref{fig:12} shows that under the joint action of convective and advective accretion flows, two shocks in case of $\lambda_0 = 1.80$ merge, which exactly corresponds to $t\approx7.6s$ burst of luminosity. When the expanding shock disappears, the accretion flow material will smoothly enter the black hole, resulting in a temporary increase in the accretion rate until a new expanding shock is born. This mechanism also makes the inner shocks more unstable. In other words, although multiple shocks can be confirmed to exist in simulations, accurate tracking and positioning still face greater challenges.

\subsection{Advection and Convection}
\label{sec:4.2}

\begin{figure}
\centering
\includegraphics[width=0.45\columnwidth]{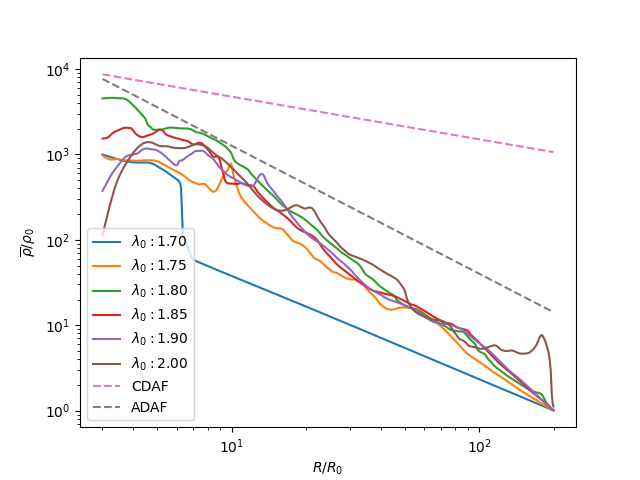}
\includegraphics[width=0.45\columnwidth]{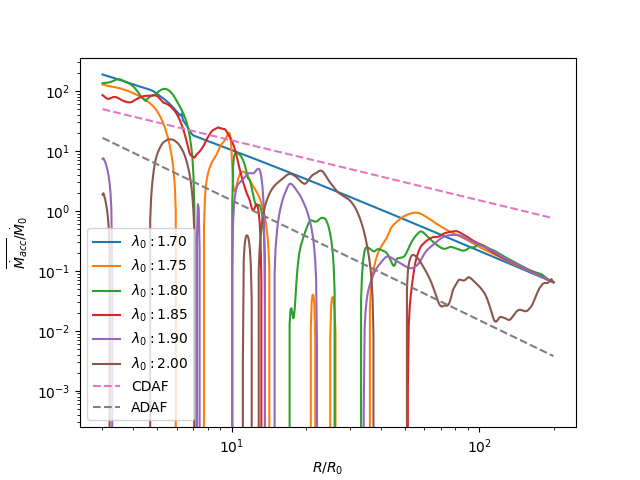}
\caption{\label{fig:22} Comparison of the dimensionless average density ($\overline{\rho}/\rho_0$, left) and average accretion rate ($\overline{\dot{M}_\text{acc}}/\dot{M}_0$, right) in the accretion region at $90000 t_0$ for $\lambda_0$= 1.70, 1.75, 1.80, 1.85, 1.90, and 2.00. The average value is obtained in the $\theta$ direction. The two dashed lines represent CDAF ($\alpha\rho=-0.5$, pink), and ADAF ($\alpha_\rho=-1.5$, gray) respectively.} 
\end{figure}

Looking at the density contours shown in Fig. \ref{fig:10}, it is apparent that even for very small $\lambda_0$ as matter flows from the outer boundary towards the inner boundary, the decreasing density of accretion flow with radius exhibits an approximate power-law like behaviour with an exponent $\alpha_\rho$. In this section, we will explore the accretion flow dynamics depicted in Fig. \ref{fig:21} by discussing the spectral index $\alpha_\rho$. Observational data for Sgr A* and M87* suggest that  $\alpha_\rho\sim-1$ \citep{2023MNRAS.521.4277R,2015MNRAS.451..588R}. Convection-dominated accretion flow (CDAF) models \citep{2000ApJ...539..798N,2000ApJ...539..809Q} posit that strong convective motion dominates the accretion process. In such flows, radiation is inefficient in carrying away the heat generated by accretion. Instead, convective motion becomes the primary mechanism for energy transfer, resulting in $\alpha_\rho\sim-0.5$. In contrast, advection-dominated accretion flow \citep[ADAFs,][]{1994ApJ...428L..13N} models propose that energy transfer is dominated by advective motion, leading to $\alpha_\rho\sim-1.5$.

Figure \ref{fig:22} displays the comparison of dimensionless average density and average accretion rate for $\lambda_0$= 1.70, 1.75, 1.80, 1.85, 1.90, and 2.00. 
It is essential to note that the two dashed lines (CDAF and ADAF) correspond to the average accretion rate $\alpha_{\dot{M}}=\alpha_\rho-0.5$ since $\dot{M} = v \cdot \rho$ with $v$ being the Keplerian velocity $\sim R^{-\frac{1}{2}}$. Both $\alpha_\rho$ and $\alpha_{\dot{M}}$ exhibit a similar, or even steeper, decline rate (i.e., $\alpha\lesssim-1.5$) compared to ADAFs, which is expected as ADAFs have larger angular momentum compared to the low-angular momentum advective flows studied in this work. The discussion focuses on $\alpha_\rho$ due to occasional distortions in $\alpha_{\dot{M}}$ caused by the logarithmic operations, leading to abrupt declines in accretion rates. By examining the region near the inner boundary ($R\lesssim10R_g$), all cases show a decline rate close to that of CDAF. Combining this with Figure \ref{fig:10}, it becomes evident that not all matter can penetrate the inner boundary and enter the black hole region. Additionally, not all matter forms a disk wind before reaching the inner boundary. This implies that a portion of the disk wind is contributed by convection near the inner boundary. This region exhibits convection-dominated motion, until encountering critical radius $R_{c}$ with a density-increasing peak. The location of $R_{c}$ increases with growing $\lambda_{0}$, possibly marking the extent of the dense region near the inner boundary. Subsequently, in most cases, the flow transitions into a stage dominated by advective motion. However, for very large angular momentum ($\lambda\geq2.00$), the decline rate in the range $50R_{g}\lesssim R\lesssim180 R_{g}$ reduces again, entering the range $-1.0\lesssim\alpha_\rho\lesssim-0.5$. This is also evident in Figure \ref{fig:21}, where a large amount of material accumulates in a massive vortex within the accretion disk. This suggests that convective motion becomes dominant again in this region. Here, due to the substantial material accumulation in the vortex, $\alpha_\rho$ becomes very large for $R\gtrsim180R_{g}$, approaching a vertical line in Figure \ref{fig:22}.

 \section{Summary}
 \label{sec:5}
 
 We conducted a series of simulations (refer to Table \ref{tb:1} for details.) using PLUTO code \cite{2007jena.confR..96M} based on the theory of Chakrabarti's transonic accretion flows \citep[e.g.,][]{1989ApJ...347..365C,1990ttaf.book.....C,1993ApJ...417..671C} and the previous numerical works by \cite{Dongsu1997Zero}. 
 
 Compared to the work of \cite{Dongsu1997Zero}, we employ spherical coordinates and focus on the first and fourth quadrants. Additionally, we determine the height of the injected flow ($\frac{H_0}{R_0}$) as per vertical equilibrium condition (Eq. \ref{eq:11}). According to our methodology, even with the parameter settings from \cite{Dongsu1997Zero}, this height should be approximately 15\% of radial distance, not their fixed value of 10\% . Using our transonic solutions (Eq. \ref{eq:12}) and assuming an injected stream with a Mach number of 5, the resulting shock positions are in the range $10-100 R_g$. This implies that our injection flow height ($\frac{H_0}{R_0}$) is around 28\% (see Table \ref{tb:1} and Fig. \ref{fig:2}). We also utilize a higher resolution and grid distribution (see Section \ref{sec:2} and Eqs. \ref{eq:14}, \ref{eq:15}). 

 We began by analyzing the direct measurements of the density profile and mass flux. We found that when there are no shocks in the system, the mass flux is very stable (e.g., when $\lambda_0 = 1.50$). However, when shocks appear, both the accretion rate ($\dot{M}_\text{acc}$) and outflow rate ($\dot{M}_\text{out}$) show fluctuations. Moreover, as the specific angular momentum ($\lambda_0$) increases, the system transitions from accretion-dominated to outflow-dominated. When $\lambda_0 \geq 1.85$, it becomes challenging for matter to fall into the black hole (see Figures \ref{fig:11}). Some theoretical solutions \citep[e.g.,][]{1989ApJ...347..365C, 1990ttaf.book.....C, 1993ApJ...417..671C} and previous simulations by \cite{Dongsu1997Zero} reported that matter struggles to fall into the black hole only when $\lambda_0 > 2.00$. In comparison, our results are somewhat different, even though we have not considered other physical mechanisms affecting the accretion flow, such as radiative pressure and viscosity. In addition, we discovered shocks within the accretion flow at $\lambda_0 = 1.75$, whereas these theoretical solutions suggest that shocks are hard to sustain when $\lambda_0 < 1.782$. This finding also contrasts with the results of \cite{Dongsu1997Zero}, who did not report shocks and their oscillations for $\lambda_0 = 1.75$ in their simulations. We speculate that this discrepancy might arise from the introduction of new sonic points when using a simulation domain spanning the first and fourth quadrants, leading to the emergence of these shocks (even if they are unstable).

 We further discuss the variation of shock positions and found that those shocks are not stable. For smaller values of $\lambda_0$ (such as 1.75), the shock oscillates around a fixed position. However, for larger values of $\lambda_0$ (such as 1.9), the oscillation is more distant from the central black hole. Given our focus on adiabatic inflow, a plausible explanation could be the advective-acoustic cycle process \cite[e.g.,][]{2006Numerical2,2008Multidimensional}. In this process, oppositely directed sound waves propagating outwards from the shock region reflect and amplify at specific locations (like boundaries, dense areas, etc.) and reconverge at the shock front. When reconverging, the misalignment in their amplitudes (i.e., unequal energies) can cause a displacement in the shock position. In this study, while we haven't delved deeply into the direct implications of the advective-acoustic cycle, we have explored the relationship between shock position variations and accretion rate (see Figure \ref{fig:11}). We observed that the evolution of the shock and accretion rate presents contrasting behavior, particularly pronounced for higher $\lambda_0$ values (like 1.9). We also noted that their variations don't occur simultaneously but maintain a time lag. This is due to the region calculating the accretion rate being at the inner boundary, with the shock being several tens to 170 $R_g$ away. The sound waves require time to propagate from the shock surface to the inner boundary, which precisely corresponds to this lag value. Intriguingly, the position of the shock typically coincides with the dense region of the accretion disk, implying a significant contribution to Bremsstrahlung radiation at this location (refer to Figure \ref{fig:8}). 

Oscillations in luminosity were observed even at $\lambda_0 = 1.70$. It's essential to understand that these oscillations predominantly stem from the intermittent shocks produced by collisions between the inflows and outflows in the funnel region, rather than from shocks inherent within the accretion disk itself (refer to Figure \ref{fig:14}). However, as $\lambda_0$ increased to values like 1.75 and 1.80, oscillatory behaviors in luminosity became evident once the system reached equilibrium, therefore, we reported luminosity oscillations with a few Hz, respectively. No discernible luminosity oscillations were found for $\lambda_0 \geq 1.85$. In these cases, the existence of multiple shocks and luminosity bursts were noted (see Figures \ref{fig:12} and \ref{fig:17} for example). Section \ref{sec:4.1} provides a detailed discussion on the occurrence of multiple shocks, and we believed the combination of multiple shocks can cause the luminosity outbursts. Simulations by \cite{2022MNRAS.514.5074O} also reported multiple shocks, finding that the inner shocks showed an opposing evolutionary trend to the outer ones; as the inner shock moved inwards, the outer one propagated outwards. 

In the cases of $\lambda_0 = 1.90$ and $1.95$, the Mach number drops rapidly after passing a strip structure that is nearly perpendicular to the equatorial plane (near $R\sim18R_g$, see Fig. \ref{fig:21}), which means that the fall of matter will be blocked here. This is the result of a strong, stable and nearly perpendicular to the equatorial plane shock. The strip structure in the Mach number map will be destroyed with $\lambda_0 \geq 2.00$ because the larger angular momentum makes it more difficult for matter to fall and easier to form a disk wind. In \cite{2019A}, it was suggested that Type C low-frequency (LFQPO), characterized by a centroid frequency $\leq$ 30 Hz, is likely induced by accretion rate fluctuations, as they are commonly observed alongside broad-band, flat-topped noise \citep{10.1093/mnras/stt1107,10.1093/mnras/292.3.679}. In Figure \ref{fig:13} and \ref{fig:15}, we further confirmed this by observing similar variations in accretion rate $\dot{M}_\text{acc}$ and the light curves, particularly pronounced at $\lambda_0 = 1.75$. \cite{2008A&A...489L41C,1996ApJ...457..805M} proposed a shock oscillation model employing a two-component accretion flow (TCAF) with varying accretion rates, leading to discontinuities at shock fronts. Further research indicates that it is these oscillations that give rise to the QPO in the X-ray flux, with the oscillation frequency inversely proportional to the infall time \citep[refer to][]{2008A&A...489L41C,1996ApJ...457..805M}, which can be approximated as:
\begin{equation}
\label{eq:19}
\nu_\text{{qpo}} = \frac{\nu_{s0}}{\mathcal{R} R_s\sqrt{R_s-1}},
\end{equation}
where $\nu_\text{qpo}$ represents the QPO central frequency, $\nu_{s0}=c/R_g$ is the inverse of the light crossing time of the black hole, $c$ is the speed of light, $R_g$ is the Schwarzschild radius, $\mathcal{R}=\rho_2/\rho_1$ is the compression ratio, and $R_s$ is the average shock location (unit of $R_g$). Taking $\lambda_0 = 1.75$ as an example (see Figure \ref{fig:10}), we have $R_s\sim 65$ and $\mathcal{R}\sim 3 - 4$, which yields $\nu_\text{qpo}\sim 3.91 - 6.52$Hz. This is in agreement with the oscillation peaks found in Figure \ref{fig:15}. This low frequency oscillation is accompanied by the accretion disk wind's direction changes, as illustrated in Figure \ref{fig:10}. During these changes, the accretion rate becomes higher, which is represented by a redder appearance in the figure. In the light curves, these moments manifest as occasional bursts (see Figure \ref{fig:13} and \ref{fig:15}). In contrast, for cases with more stable accretion flows with either higher or lower specific angular momentum, such bursts are not as prominent, as there is no change in the accretion disk's direction in these cases. In addition, $\lambda_0 = 1.70$ and 1.75 also show oscillation peaks in the high frequency band. One possible mechanism involves acoustic resonances due to pressure and accretion rate interactions \citep[see][section 5.6 and references therein]{2019A} and classical models related to oscillations, e.g., relativistic resonance models \citep{2004ApJ...603L..93L} and MHD model \citep[e.g.][]{shi.chang.zhang.li.2014, 2010ApJ...714.1227S, 2018MNRAS.479.5049S}. In such cases, $\nu_\text{qpo} \sim \frac{2\pi c_s}{r_\text{tl}}$, where $r_\text{tl}$ denotes the truncated radius. In our scenario, this results in frequencies ranging from several tens to several hundreds of Hz, precisely matching the less distinct high-frequency peaks observed in the power spectrum.

In summary, from simulations of adiabatic low-angular momentum advective flow, we observed shocks for $\lambda_0 \geq 1.75$ and significant luminosity oscillations within the range of $1.70 \leq \lambda_0 \leq 1.80$. It's worth noting that our findings slightly differ from those of \cite{Dongsu1997Zero} since they didn't detect shocks at $\lambda_0 \leq 1.75$ and theoretical prediction. This discrepancy might arise from considering the fourth quadrant in 2D simulations, where we speculate that more transonic sonic points could exist. Considering known observations, we identified that oscillations in the range of several to several tens of Hz could be attributed to shock oscillations and lower-frequency (0.1-1 Hz) luminosity oscillations and outbursts caused by changes in the disk wind direction. Several sources, such as GX 339-4 and XTE J1859+226, have reported QPOs at several Hz in multiple observations \citep{2019A}. It's intriguing to note that GX 339-4 exhibited an outburst in 2021, coinciding with its observed QPOs \citep{2023MNRAS.526.4718M}.

Our dimensionless calculations allow us to extend the applicability of our model to Sgr A*. The resulting frequency range is $10^{-6}-10^{-5}$ Hz, which agrees well with some observational data concerning Sgr A* \cite[e.g.,][]{2013The,2013A,2014THE,2015Fifteen}. However, it's important to note that our model is simplified and doesn't account for magnetic fields or full general relativistic treatment. Recent simulations by \cite{2019PASJ...71...49O} and \cite{2022MNRAS.514.5074O} have shown that magnetic fields can induce shocks at lower $\lambda_0$ values, and the results of \cite{2018Magnetized} further support this scenario. 

\section*{Acknowledgements}
This work is supported by the National Natural Science Foundation of China under grant No. 12073021. We would like to thank Prof. Toru Okuda for helpful discussions regarding this work. The numerical simulations were conducted on the Yunnan University Astronomy Supercomputer (YUNAS) and analyzed by open-source tools \textit{ShockFinder} (see \url{https://www.github.com/wacmkxiaoyi/shockfinder}).

\bibliographystyle{raa}

\appendix

\section{Simulation run with $\gamma=5/3$}
\begin{figure}[H]
\centering
\includegraphics[width=0.8\columnwidth]{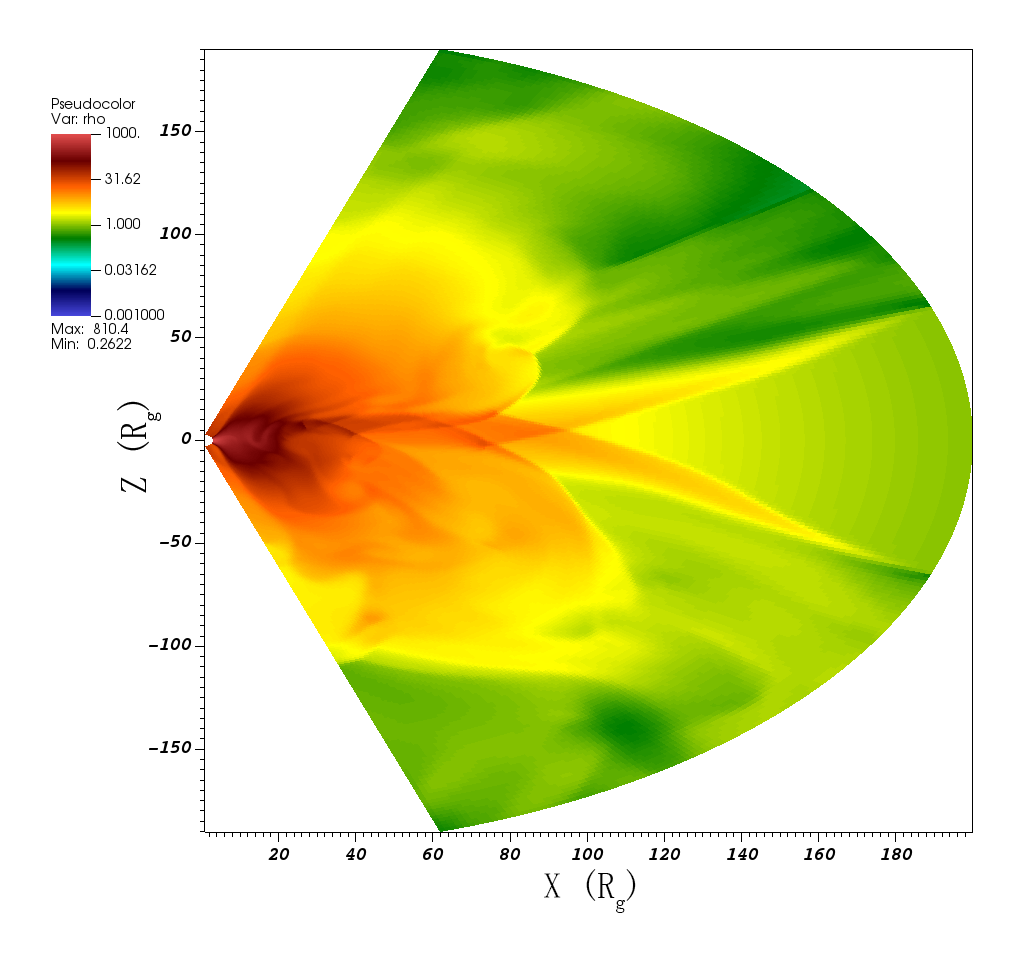}
\caption{\label{fig:apx1} The density profile of the simulation with $\lambda_0 = 1.75$ with $\gamma = 5/3$ in $90000t_0$.}
\end{figure}

We also ran the simulation case $\lambda_0 = 1.75$ with $\gamma = 5/3$, and we found that the shock evolution is similar to that of $\gamma = 4/3$. The shock oscillates around the equatorial plane, even though the high-density region is not limited to the vicinity of the black hole for $\gamma = 5/3$, because the higher adiabatic index will make the gas require a higher pressure to compress (i.e., $pV^\gamma$ is a constant). Therefore, our simulation results are also relevant to different $\gamma$.

\end{document}